\documentclass[11pt,preprint]{aastex} \usepackage{graphicx}

\def\rpro{\mbox{$r$-process}}
\def\spro{\mbox{$s$-process}}
\def\teff{\mbox{T$_{\rm eff}$}}
\def\logg{\mbox{log~{\it g}}}
\def\vmicro{\mbox{$\xi_{\rm t}$}}

\shorttitle{Chemical abundances in Omega Centauri}
\shortauthors{Marino et al.}
\usepackage{ulem}

\begin{document}

\title{Sodium-oxygen anticorrelation and neutron capture elements in Omega Centauri stellar populations
          \footnote{Based  on data collected at the European Southern
    Observatory with the VLT-UT2, Paranal, Chile.       }}

\author{
   A.\ F.\ Marino\altaffilmark{1, 2},
	  A.\ P.\ Milone
	  \altaffilmark{3, 2},
          G.\ Piotto
	  \altaffilmark{2},
	  S.\ Villanova
          \altaffilmark{4},
	  R.\ Gratton
	  \altaffilmark{5},
	  F.\ D'Antona
	  \altaffilmark{6},
          J.\ Anderson
          \altaffilmark{7},
          L.\ R.\ Bedin
          \altaffilmark{7},
          A.\ Bellini
          \altaffilmark{2, 7},
          S.\ Cassisi
          \altaffilmark{8},
          D.\ Geisler
	  \altaffilmark{4},
	  A.\ Renzini
	  \altaffilmark{5},
	  and
	  M.\ Zoccali
	  \altaffilmark{9}
}

\altaffiltext{1}{Max-Planck-Institut f{\"u}r Astrophysik, Postfach 1317, D--85741 Garching b. M{\"u}nchen, Germany; amarino@mpa-garching.mpg.de}

\altaffiltext{2}{Department of Astronomy, University of Padova, Vicolo dell'osservatorio 3, 35122, Padova, Italy; anna.marino@unipd.it}

\altaffiltext{3}{Instituto de Astrof\'isica de Canarias, La Laguna, Tenerife,
  Spain}

\altaffiltext{4}{Departamento de Astronom\'ia, Universidad de Concepci\'on, Casilla 160-C, Concepci\'on, Chile}

\altaffiltext{5}{INAF-Osservatorio Astronomico di Padova, Vicolo dell'Osservatorio 5, 35122 Padova, Italy}

\altaffiltext{6}{INAF-Osservatorio Astronomico di Roma, via Frascati 33, I-00040 Monteporzio, Italy}

\altaffiltext{7}{Space Telescope Science Institute, 3700 San Martin Drive, Baltimore, MD 21218, USA}

\altaffiltext{8}{INAF-Osservatorio Astronomico di Teramo, Via M. Maggini, 64100 Teramo, Italy}

\altaffiltext{9}{P. Universidad Cat\'olica de Chile, Departamento de Astronom\'ia y Astrof\'isica, Casilla 306, Santiago 22, Chile}

\begin{abstract}
Omega Centauri is no longer the only globular cluster  
known to contain multiple stellar populations, yet it 
remains the most puzzling.
Due to the extreme way in which the multiple stellar population phenomenon 
manifests in this cluster, it has been 
suggested that it may be the remnant 
of a larger stellar system. In this work, we present a spectroscopic 
investigation of the stellar populations hosted in the globular cluster
$\omega$~Centauri  
to shed light on its, still puzzling, chemical enrichment history. With this 
aim we used FLAMES+GIRAFFE@VLT to observe 300 stars distributed along the 
multimodal red giant branch of this cluster, sampling with good statistics 
the stellar populations of different metallicities. We determined chemical 
abundances for Fe, Na, O, and $n$-capture elements Ba and La. We confirm that 
$\omega$~Centauri exhibits large star-to-star variations in iron with 
[\rm Fe/H] ranging from $\sim -$2.0 to $\sim -$0.7 dex. Barium and lanthanum 
abundances of metal poor stars are correlated with iron, up to [Fe/H]$\sim
-$1.5,  
while they are almost constant (or at least have only a moderate increase) in 
the more metal rich populations. There is an extended Na-O anticorrelation for 
stars with [Fe/H]$\lesssim -1.3$ while more metal rich stars are almost all 
Na-rich. Sodium was found to midly increase with iron over all the metallicity
range.

\end{abstract}

\keywords{globular clusters: individual (NGC 5139)
            --- stellar populations -- chemical abundances }

\section{Introduction}
\label{introduction}
Omega Centauri ($\omega$~Cen) is one of the most studied and enigmatic stellar
systems in our Galaxy. 
It hosts the most complex multiplicity of stellar populations 
observed in globular clusters (GC), with an anomalous large spread in [\rm
  Fe/H], ranging from  
$\sim-$2.0 to $\sim-$0.6 dex (Norris, Freeman \& Mighell\ 1996; Suntzeff
  \& Kraft 1996; Lee et al.\ 1999;  Pancino et al.\ 2000; Sollima et
  al.\ 2005a; Johnson et al.\ 2008; Johnson \& Pilachowski 2010). 

Evidence for large variations in the light elements among $\omega$~Cen stars
at a given [Fe/H] was initially detected by Persson et al.\ (1980) in the form
of a wide spread in CO absorption from infrared photometry. Brown \&
Wallerstein (1993) and Norris and Da Costa (1995) saw a similar variation in
Na, Al and CNO elements. More recently, the variations in O, Na, and Al have
been confirmed  as well by spectroscopic studies conducted on a large sample
of red giants (see Johnson \& Pilachowski 2010, and reference therein).  

Abundance ratios of slow neutron-capture process (\spro)\ elements
  relative to iron  exhibit large star-to-star variations, with a significant
  number of stars showing enhancements over the entire metallicity range
  (Johnson \& Pilachowski 2010 for red giants, and Stanford et al.\ 2007 for
  sub giant and main sequence stars). In addition, the ratio of
  \spro\ abundances to Fe increases with [Fe/H] for low [Fe/H], but becomes
  constant at higher [Fe/H] (Norris \& Da Costa 1995; Smith, Cunha \&
  Lambert\ 1995; Stanford, Da Costa \& Norris\ 2010; Johnson \&  Pilachowski
  2010).  

The complexity of multiple stellar populations in  $\omega$~Cen can also be
seen in the multiple red-giant branches (RGB) and  
sub-giant branches (SGB) (Lee et al.\ 1999; Pancino et al.\ 2000; Ferraro et
al.\ 2004; Sollima et al.\ 2005b; Bellini et al.\ 2009). 
Anderson\ (1997) and Bedin et al.\ (2004) clearly detected that the main
sequence (MS)  
splits into two main branches. Contrary to any expectation from standard
stellar evolutionary  
models, the more metal poor stars populate a redder MS (rMS), and stars at
intermediate  
metallicity a bluer one (bMS; Piotto et al.\ 2005). To date, the only
explanation we have  
to account for these observations is to assume a high overabundance of He for
the bMS stars  
(Bedin et al.\ 2004; Norris 2004; Piotto et al.\ 2005). A third, less
populated MS (MSa) has 
been discovered by Bedin et al.\ (2004) on the red side of the rMS and it has
been  
associated with the most metal rich stars (Villanova et al.\ 2007). As well as
bMS stars,  
the colors and the metallicity of the MSa are consistent with being 
 populated by He rich stars (Norris \ 2004, Bellini et al.\ 2010).

All the observational data suggest that $\omega$~Cen has experienced a complex 
evolution with a still puzzling chemical enrichment history. As a matter of
fact, we still do not know how such a large iron spread could have originated
in a GC.  For this reason, it has been suggested that $\omega$~Cen may not
really be a GC, but may rather be the remnant of a larger system (Bekki \&
Norris 2006, and references therein). 

Recent spectroscopic detections of intrinsic variations in [\rm Fe/H] 
in the GCs NGC~6656 (M22, Marino et al.\ 2009), NGC~6715 (M54, Carretta at
al.\ 2010), and Terzan~5 (Ferraro et al.\ 2009) suggest that $\omega$~Cen may
not be a unique case, but rather may simply be an extreme example
of a self-enriching cluster. 
While nearly all 
GCs studied to date show no significant variations in their $n$-capture
elements (Ivans et al.\ 1999, 2001; James et al.\ 2004; Yong et
  al.\ 2008; Marino et al.\ 2008; D'Orazi et al.\ 2010), a bimodal  
distribution in the abundance of these elements has been detected in M22
(Marino et al.\ 2009),  
and NGC 1851 (Yong \& Grundahl\ 2008) and has been associated with the bimodal
SGB of these GCs. 
In particular, it appears that the nucleosynthetic enrichment processes in
$\omega$~Cen and M22 were very similar (Da Costa \& Marino 2010). 

Omega Centauri is not completely dissimilar from mono-metallic globular 
clusters.  Like all GCs studied so far, $\omega$~Cen stars show 
variations in light-element abundances, such as the well-known
pattern of Na-O anticorrelation  
(Norris \& Da Costa \ 1995, hereafter NDC95). It is now largely accepted that
sodium and oxygen  
variations indicate the presence of different stellar populations in GCs, with
a second  
generation enhanced in Na and depleted in O due to material that has undergone
the entire  
CNO cycle in first generation stars (Ventura et al.\ 2001; Gratton et
al.\ 2001; Gratton, Sneden \& Carretta 2004). 
Proposed polluters to account for the Na-O anticorrelation observed in GCs are
either AGB stars (D'Antona \& Caloi 2004), fast rotating massive stars
(Decressin et al.\ 2007), massive binaries (De Mink et al.\ 2009). At odds
with these scenarios, an alternative model, recently proposed by Marcolini et
al.\ (2009), considers as the first generation those stars that are O-depleted
and Na-enhanced, while O-rich and Na-depleted stars should be formed
successively after the self-pollution from SNe~II.  
As observed for some GCs, the groups of stars with different Na and O can
account for the 
distinct RGB sequences observed in the CMD (Marino et al.\ 2008, Milone et
al.\ 2010). 
Hence the position of a star in the Na-O plane gives an
important indication as to what population it belongs to.
According to this, by homogeneously analyzing the Na-O anticorrelation 
in the largest sample of GC stars studied so far,  
Carretta et al.\ (2009), on the basis of their position on the Na-O plane,
divided stars into primordial, intermediate, and extreme populations,
confirming that the Na-O anticorrelation could be a useful tool to study
multiple stellar population phenomenon in GCs. 

In this article, we analyze a large sample of GIRAFFE@VLT  spectra and
determine chemical abundances of iron, sodium, oxygen, barium, lanthanum for
hundreds of RGB stars in $\omega$~Cen. 
Thanks to this large sample of stars, we are also able to
study the Na-O anticorrelation separately for stars in 
different metallicity populations.

The layout of this paper is as follows: Section 2 is a brief overview of the
data set and the spectroscopic analysis; our results, and a comparison with
previous studies, are presented in Section 3; Section 4 is a brief summary and
discussion of 
the results obtained in this study.

\newpage
\section{Observations and data analysis}
\label{obs}

%
The data set consists of a large number of FLAMES/GIRAFFE spectra [proposals:
  071.D-0217(A), 082.D-0424(A)] taken with the four set-ups HR09B, HR11, HR13,
and HR15, which cover a spectral coverage of 5095-5404, 5597-5840, 6120-6405,
6607-6965 \AA, respectively, with a resolution $R$ $\sim$ 20,000-25,000. 
The sample is composed by 300 red giants with a $V\leq$15. The typical S/N of
the final combined spectra is $\sim$60-100, with values of S/N$\sim$130-160 in
the spectral region around the O line at 6300 \AA. 

Data reduction involved bias-subtraction, flat-field correction, 
wavelength-calibration, sky-subtraction, and spectral rectification.
To do this, we used the dedicated pipeline BLDRS v0.5.3, written at the Geneva
Observatory (see {\sf http://girbld-rs.sourceforge.net}). 
Chemical abundances were derived from a local thermodynamic
equilibrium (LTE) analysis by using the spectral analysis code MOOG
(Sneden 1973).

Atmospheric parameters (\teff, \logg, \vmicro, metallicity) were estimated as
follows. Effective temperatures (\teff) were obtained via empirical relations
of the $(V-K)$ colors with temperatures using the 
calibrations by Alonso, Arribas \& Mart\'inez-Roger (1999, 2001).
We used $V$ magnitudes from Bellini et al.\ (2009), $K$ magnitudes
retrieved from the Point Source Catalog of 2MASS (Skrutskie et
al.\ 2006) and then transformed to the Carlos S\'anchez Telescope 
photometric system, as in Alonso et al.\ (1999).
As noted by Bellini et al.\ (2009), $\omega$~Cen is not
significantly affected by differential reddening. This allowed us to apply
a uniform reddening correction to all stars, by correcting our
final $(V-K)$ colors for interstellar extinction using a $E(B-V)=$0.12
(Harris 1996) and $E(V-K)/E(B-V)=$2.70 (McCall 2004).
At the end, as a test we verified that our adopted photometric temperatures
satisfied excitation equilibrium by plotting the Fe~I abundances versus the
excitation potentials and verifying that they did not reveal any trends. 

Surface gravities in terms of \logg\  were obtained from the apparent
magnitudes, 
the above effective temperatures, assuming a mass M~$=~0.80~{\rm
  {M_{\odot}}}$, bolometric corrections from Alonso et al.\ (1999), and
a distance modulus of $(m-M)_{V}=$13.69 (Cassisi et al.\ 2009).
Microturbolent velocities were determined by removing trends in the relation
between abundances from neutral Fe lines and reduced equivalent widths (see
Magain 1984). 
Final metallicities were then obtained interactively by interpolating
the Kurucz (1993) grid of model atmospheres with \teff,\ \logg,\ and \vmicro\ fixed to find the [Fe/H] that best matched the Fe~I lines.

Abundances from Fe lines were derived from equivalent-width (EW)
measurements, but we derived the abundances for the other elements
from spectral synthesis.  The sodium content was determined from the
doublet
at $\sim$6150 \AA. The oxygen abundances were derived from the forbidden line
at 6300 \AA, after the spectrum was cleaned from telluric contamination, using
a synthetic spectrum covering the spectral region around the O line and taking
line positions and EWs for the atmospheric lines of the Sun provided by Moore,
Minnaert \& Houtgast (1966). Barium was measured from the blended Ba~II line
at 6141 \AA, and lanthanum from the two La~II lines at 6262 \AA \ and 6390
\AA. The hyperfine splitting was taken into account for the La spectral lines
by using data from Lawler, Bonvallet \& Sneden (2001).  
Marino et al.\ (2008) provide more details on the linelist.

An error analysis was performed by varying the temperature, gravity,
  metallicity, microturbolence, and repeating the abundance analysis for stars
  representative of all the observed range in [Fe/H]. The parameters were
  varied by $\Delta$(\teff)=$\pm$50 K, $\Delta$(\logg)=$\pm$0.20 dex,
  $\Delta$(\vmicro)=$\pm$0.10 km/s and metallicity by $\pm$0.10 dex, which
  are typical internal uncertainties for stars with similar temperatures as
  ours and analyzed with spectra of similar quality (see Carretta et
  al.\ 2009). We redetermined the abundances by changing only one atmospheric
  parameter at a time for several stars, to see
  how abundances are sensitive to these variations in the atmospheric models. 
  Changes in \teff\ mostly affect the abundances derived from the subordinate 
  ionization state transitions, i.e.\ Na~I and Fe~I abundances.
  On the contrary, abundances derived from the dominant ionization state
  transitions, i.e.\ singly ionized Ba, La, and Fe, and neutral O, are
  more sensitive to variations in gravity and metallicity. 
  Errors introduced by changes in \vmicro\ increases with increasing metallicity.   
  Assuming that the parameter uncertainties are uncorrelated, the resulting
  errors in chemical abundances were calculated by summing in quadrature with
  the contributions to errors given by each parameter. 
An additional source of internal errors for Fe chemical abundances is the
uncertainty in the EW measurements, which we quantified to be $\sim$5
m\AA.\ We estimated this quantity by comparing EW measurements for the same
spectral lines of pairs of stars with similar atmospheric parameters. In the
case of Fe~I this contribution is small since a large number of transitions
(typically $\simeq$ 20-30) are available. We estimate the
uncertainty in the [Fe/H] abundance from our EW measurements by assuming that
the error will be $\sigma_{\rm EW}$/$\sqrt{N_{\rm lines}-1}$, where
$\sigma_{\rm EW}$ is the dispersion in abundances and $N_{\rm lines}$ the
number of spectral lines for a given specie. On average we estimated the EW
error contribution to be $\sim$0.02~dex for Fe~I, and $\sim$0.06~dex for
Fe~II.  
We summed these [Fe/H] uncertainties in quadrature to the
uncertainties introduced by atmospheric parameters and found
that typical internal errors in iron abundances are of the order of 0.05-0.07
dex for Fe~I, and of $\sim$0.10 dex in the case of Fe~II.  
For elements for which the abundances were derived from
spectral synthesis, internal scatter factors
are introduced by the errors in fitting the observed 
spectrum to the model, which included the fit to the
continuum.
For our spectra at relatively high S/N, this translates in typical uncertainties of
$\sim$0.05 dex. 
The total uncertainty, which includes the error introduced by the
uncertainties in the model atmosphere, obtained by repeating the spectral synthesis
for different atmospheric parameters, as explained above, were estimated
to be of the order of $\sim$0.10-0.15 dex in Ba and La, and of $\sim$0.10 in Na and O.

\section{Results}
\label{gir}
In this Section we present our results for the five analyzed elements, that
are  of particular interest in the study of multiple stellar populations in
$\omega$~Cen. 
The abundance ratios measured in this work, together with adopted atmospheric parameters, and $V$ magnitudes are listed in Tab.\ref{list}, fully available in electronic form at the CDS ({\sf http://cdsweb.u-strasbg.fr}).

\begin{figure}[!ht]
\centering
\plotone{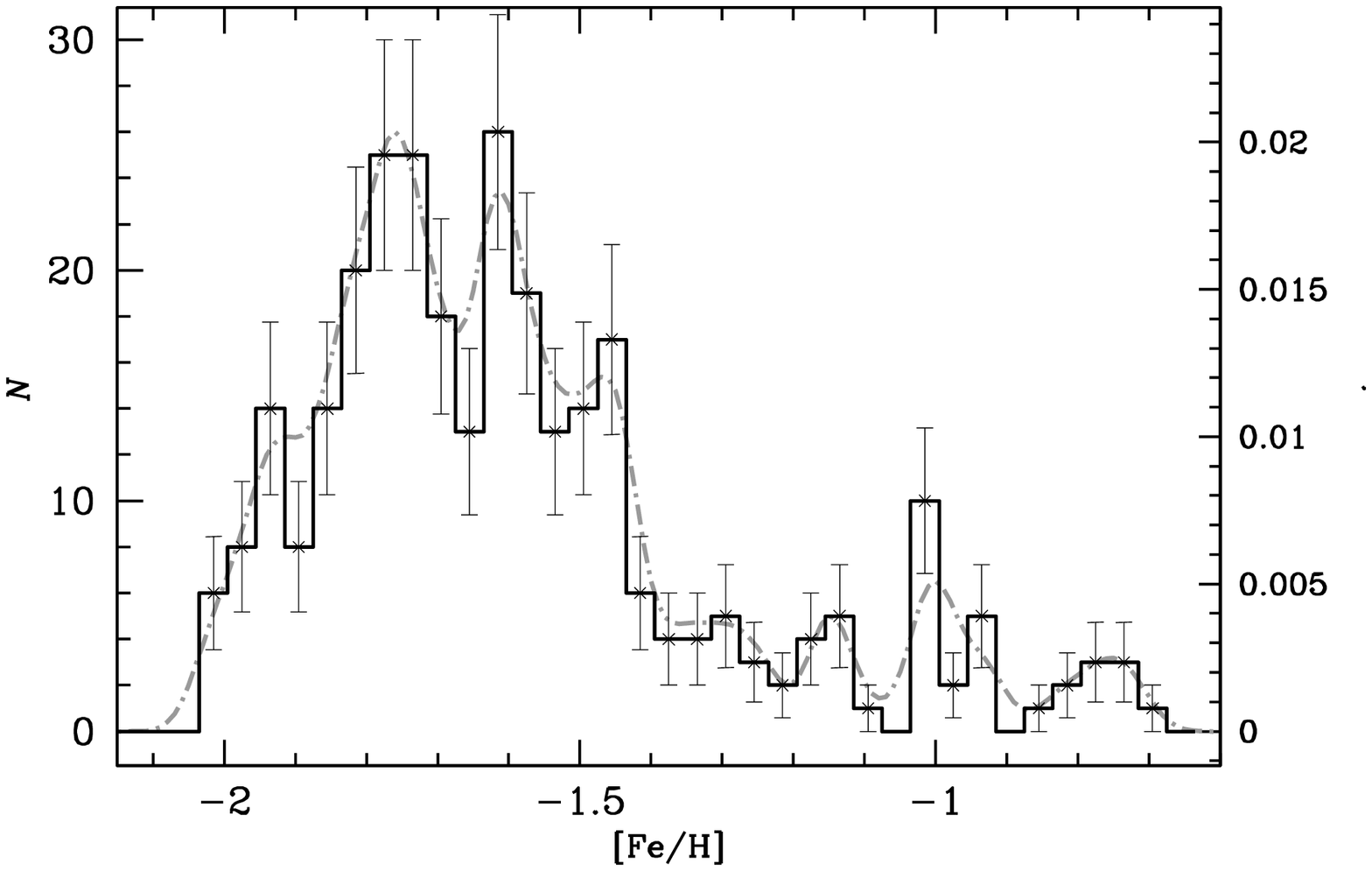}
\caption{Histogram of [Fe/H] distribution. The superimposed dashed
  line represents the normalized kernel density distribution, the
  error bars are the Poisson errors.}
\label{kern}
\end{figure}

\begin{enumerate}
{\item \bf Iron.}
We found that in $\omega$~Cen [Fe/H] ranges from $\sim -2.0$ to $\sim -$0.7
dex and follows the 
histogram distribution shown in Fig.~\ref{kern}, where error bars represent the Poisson errors.
To estimate the significance of the revealed peaks, we smoothed
the observed iron distribution with
the normalized kernel density distribution $K=\sum_{i=1}^{N}
e^{\frac{-(x-x(i))^{2}}{2 \sigma^{2}}}$ (gray dot-dashed line superimposed to
the observed histogram distribution)
where:
{\it N} is the number of stars for which the iron abundance has been
determined in this paper, 
{\it x(i)} is the [Fe/H] measured for each star,
{\it x}=[Fe/H], and $\sigma$ has been taken equal to 0.05 dex that is similar
to the internal observational error associated with [Fe/H]. 

The iron distribution represented in Fig.~\ref{kern} is in agreement
with previous studies, which have shown that stars in $\omega$~Cen span a
wide interval in metallicity with almost no stars with [Fe/H]$< -$2.0, the
majority 
having [Fe/H]$\sim -$1.7 and the remaining stars forming a tail at
higher metallicities up to [Fe/H]$\sim -$0.7 (see Johnson et al.\ 2008; 2009;
Johnson \& Pilachowsky 2010, hereafter J08/09/10, and references therein). 

From a photometric analysis of the multimodal RGB, Sollima et al.\ (2005b)
resolved five metallicity groups, going from the dominant RGB metal poor group
with [Fe/H]$\sim -$1.7 to the more metal rich stars peaked at [Fe/H]$\sim
-$0.7. 
A multimodal peak distribution was also found in Str\"omgren photometry by
Calamida et al.\ (2009). 
The $\omega$~Cen metallicity distribution obtained from spectroscopic studies
obviously depends on the sample statistics and on selection biases.
Spectroscopic studies based on medium resolution data for large samples of
$\omega$~Cen stars include:
Sollima et al.\ (2005a, 250 SGB stars), Villanova et al.\ (2007, 80 SGB
stars), and J08/09/10 (800 RGB stars in total). 
Sollima et al.\ (2005a) and J08/09/10
identified  four groups of stars on the SGB and the RGB, respectively, peaked
at values of [Fe/H]$\sim -$1.75, $-$1.45, $-$1.05, and $-$0.75 (values from
J09), in agreement with the distributions obtained from RGB photometry
(Sollima et al.\ 2005b). A similar distribution was found by Villanova and
coworkers, but, possibly due to their lower statistics, they did not find
stars in the more metal reach peak of Sollima et al.\ (2005a, b) associated
with the fainter SGB (SGB-a). 

An inspection of Fig.~\ref{kern} suggests that our metallicity distribution is
consistent with the presence of multiple peaks corresponding to
[Fe/H]=$-$1.75, $-$1.60, $-$1.45, $-$1.00, a broad distribution of stars
extending between $-$1.40 and $-$1.00, and a tail of metal rich stars reaching
values of [Fe/H]$\sim -$0.70. 
Interestingly, we also note a tail of stars at low metallicity down to
[Fe/H]$\sim -$1.95, in agreement with calcium triplet low-resolution studies
(Norris, Freeman, \& Mighell 1996; Suntzeff \& Kraft 1996) that have shown
that there are a few RGB stars with [Fe/H] $< -$1.80. 
However, we note that, because of observational errors, particular caution
should be adopted in the definition of multiple Fe peaks, and in establishing
the presence of discrete populations rather than a Fe gradient in agreement
with a prolonged star formation. 

Due to the target selection biases introduced by the necessity of having
a sample representative of the entire range in metallicities, and to
the presence of a radial metallicity gradient (J08 and references therein),
our metallicity distribution cannot be considered to be representative of the
distribution valid for the entire stellar populations in $\omega$~Cen. 

{\item \bf Neutron capture elements.}
The abundances of Ba and La are plotted as a function of iron content in the
panels ($a$) and ($b$) of Fig.~\ref{ONaLa}. 
Both $n$-capture elements show a trend with iron.
The abundance ratios of Ba and La increase with [Fe/H]
from the most metal poor stars to [Fe/H]$\sim-$1.5 dex,
while at higher metallicities the slope is is essentially flat.
At high metallicities, the [La/Fe] shows less scatter than [Ba/Fe] probably
due to an increase in scatter for Ba when the abundance gets high and the
lines get too strong for an accurate abundance measurement. 

It should be noted that stars with [Fe/H]$\gtrsim -$1.5 belong to the
metal intermediate RGB and RGB-a stars defined by Pancino et al.\ (2000).
Radial distribution studies provide evidence that the blue MS and the 
  metal intermediate RGB population are part of the same group of stars
(Sollima et al.\ 2005b, Bellini et al.\ 2009), while RGB-a stars, which
correspond to the progeny of 
MSa, exhibit a radial trend similar to the one observed for metal intermediate
stars. 
From these considerations we could tentatively associate metal-poor/He-normal
stars in $\omega$~Cen with stars in the metallicity range  
where La and Ba increase rapidly with iron, while metal-intermediate/He-rich
stars (and maybe RGB-a stars) could correspond to the 
stars exhibiting a flatter trend of Ba and La with metallicity. 

\item{\bf Sodium}
Our abundances show a large star-to-star Na variation, with [Na/Fe]
going from less than $\sim -$0.3 in some of the most metal poor stars,
to $\sim$1.2 in the metal rich populations.
As shown in panel ($d$) of Fig.~\ref{ONaLa}, sodium grows with iron but
there are significant variations in [Na/Fe] among stars with the same [Fe/H].
Since we can compare stars with similar atmospheric parameters, we can
safely assume that the observed Na variations are not due to
departures from LTE. Non-LTE corrections have not been applied to our Na
abundances because no standard values are provided in the
literature. However, the Na transitions used here, for our metallicity
range, typically have corrections $\leq$0.2 dex (e.g., Gratton et al.\
1999; Gehren et al.\ 2004).

\item{\bf Oxygen}
The oxygen abundance ratio to iron ranges from $\sim -$0.6 to $\sim$0.9 dex.
At variance with the other elements studied in this paper, 
there is no clear correlation with iron: a large spread is
present at all metallicities.

A visual inspection of some spectra reinforces the impression that 
some stars that have the same iron abundances, strongly differ in their
oxygen and sodium content.
As an example, in Fig.~\ref{spectra}  we show spectra of two pairs of stars with very
similar stellar atmospheric parameters for which we measured a similar iron
abundance but different oxygen and sodium. 
At variance with iron lines, the line depths of
sodium (upper panel) and oxygen (lower panel) spectral lines differ significantly and, because of the
similarity in their atmospheric parameters, must be indicative of
an intrinsic difference in these elements.
In the right panels of Fig.~\ref{spectra} we show the position of the two
selected pairs of stars on the CMD (upper panel) and on the Na-O plane (lower
panel): while each pair occupies a similar position on the CMD, the position
on the Na-O plane suggests that the two stars in each pair belong to
different stellar populations, given their different Na and O content. 
\end{enumerate}

\subsection{Comparison with literature}

In this Section we compare our atmospheric parameters and measured chemical
abundances 
with previous analysis on $\omega$~Cen giants based on moderate and
high resolution data.

Previous studies that used high-resoultion spectra 
or had large samples of RGB stars are:
NDC95 (40 stars, $R$=38,000),
J08/09/10 ($\sim$900 stars in total, $R$=13,000-18,000), Smith et al.\ (2000,
10 stars, $R$=35,000), Pancino et al.\ (2002, 6 stars, $R$=45,000), Brown et
al.\ (1991, 18 stars, $R$=17,000). 

In Fig.~\ref{bellini} we compare our values for \teff,\ \logg,\
[Fe/H], and \vmicro\ with those of previous studies for stars in common with our sample.
Data retrieved from different sources are represented with different
symbols as indicated in the inset in the upper left panel.
In addition to the works mentioned above, with gray dots, we represent
metallicities from Suntzeff \& Kraft (1996) derived from the Ca~II infrared
triplet and calibrated on the stars in common with NDC95. 
In all the panels we quoted the mean difference between our values and the
literature ones, and the scatter of the points around the linear least
squared fit with literature data. The straight line represents the perfect agreement. 

An inspection of this figure suggests that a systematic difference of
$\sim$0.15 dex in [Fe/H] is present between this study and the literature, our
[Fe/H] is lower with a scatter around a linear least squared fit of
$\sim$0.13 dex. 
Systematic differences are also present in the effective temperatures (left
upper panel) with our temperatures on average lower by $\sim$70 K with respect
to literature studies. 
As it can be seen from the figure, agreement is quite good for the surface
gravity estimates, while there is a larger scatter for the microturbulence
estimates. 

It is not trivial to track down the cause of the offsets
observed in the
[Fe/H] and \teff\ values, since
different techniques have been used by different authors to derive temperatures.
As fully explained in Section~\ref{obs}, we derived temperatures from the \teff-$(V-K)$ calibration from Alonso et al.\ (1999).
A similar method based on colors-\teff\ calibrations was used by both NDC95 and J08/09/10. 
As an example, J09 averaged the \teff\ values obtained from the
color-temperature relations of Alonso et al.\ (1999, 2001) and Ram\'irez \&
Mel\'endez (2005) for $(V-J)$, $(V-H)$, and $(V-K)$ color indices, while J08
used only the $(V-K)$ colors and then adjusted temperatures to satisfy the
excitation equilibrium for the Fe~I lines. 
Since we used the same $K$ magnitudes retrieved from the 2MASS
catalog as J08/09, possible differences in the $V$ magnitudes coming
from different photometries could explain
the differences that we observe with these authors.
In Fig.~\ref{fotometrie} (upper panel), we compare $V$ magnitudes from the WFI
photometry used in this study (Bellini et al.\ 2009, B09) with those
from Van Leewen et al.\ (2000, VL00) used by J08/09.
In the lower panels the two $V$-($B-V$) CMD are shown, with the stars studied in this work indicated by red symbols.
The comparison between the two photometries clearly shows that
on average  $V_{B09}-V_{VL00}\sim -$ 0.15.
These differences help to explain why our temperatures are systematically
lower by $\sim$50-100 K, and hence contribute to produce a systematic
difference in the measured metallicities. 
Our final choice is to use the $V$ magnitude from B09 considered to be more
reliable because these values are corrected for sky concentration effects and
calibrated on Stetson photometric secondary standards (Stetson 2000, 2005). 
We concluded that even if systematic differences in the metallicities are
present with previous studies, they will not impact our investigation, since
we are mainly interested in the star-to-star variations, which are
differential measurements. 

In Fig.~\ref{confrontoEL} we compare the elemental abundance ratios obtained
in this work with those obtained in the other studies. As it can be seen from
the figure, 
[O/Fe], [Na/Fe], and [La/Fe] abundances are in reasonable agreement with literature
values, indicating that these abundance ratios are less affected by systematic
differences in the atmospheric parameters. Larger systematic discrepancies are
present for the [Ba/Fe] abundances, for which however literature values on RGB
are available only in NDC95 and Smith et al.\ (2000).

\subsection{Sodium oxygen anticorrelation}

In this Section we analyze the Na-O anticorrelation
and attempt to further analyze the behaviour
of sodium and oxygen in stars with different iron.
Specifically, we analyzed the 253 out of 300 stars, for which we were able to derive reliable Na and O contents.
In some cases, we were able to give only an upper limit for the Na and O abundances.

On the CMD of $\omega$~Cen at least 5 distinct sequences can be
  recognized, with hints for the possible existence of other, less populous
  components (Sollima et al.\ 2005b; Bellini et al.\ 2010). If we hope to
  understand the sequence of events that led to the formation of this puzzling
  cluster and its present state, it will be critical to determine whether
  spreads and correlations in abundance patters are present only between
  sub-populations, or whether they are present within the sub-populations
  themselves.  
Indeed, if sub-populations are found to be chemically homogeneous this would
indicate that star-formation proceeded as a series of bursts, interleaved by
inactive periods during which the ejecta of intermediate mass stars were
accumulating, 
prior to a successive burst. On the other hand, if each sub-population was to
exhibit abundance spreads, then accumulation of stellar ejecta,
hypothetical dilution with pristine material, and star formation had
to proceed simultaneously. 
We know that the different metallicity of the rMS and bMS imply that the
helium abundance has at least two discrete values, as opposed to a continuous
distribution, and therefore it would be difficult to envisage a process that
would lead to a large spread in oxygen without producing a spread also in
helium.  

One first difficulty in attempting this approach is encountered when
trying to identify each of the sub-populations on the CMD. This is more
easy on high S/N Hubble Space Telescope (HST) images (see, e.g. 
the CMD from high precision Wide Field Camera 3 (WFC3)
photometry in
Bellini et al.\ 2010). However, HST data cover relatively small, and crowded, areas of the cluster,
much smaller than the 25 arcmin diameter field of view of the FLAMES
facility that we have used. Therefore, we have been forced to select our targets on 
a CMD based on wide field images taken at the ESO/MPI 2.2m telescope, 
where the distinction of the various RGBs is not as clean (see bottom
panel of Fig.~\ref{nao_new}). Therefore, in this paper, we 
separated the different RGB sub-populations
only on the basis of their iron abundance.

We also note that, unlike other clusters with prominent multiple populations
(e.g., NGC~2808), $\omega$~Cen exhibits the very wide and complex distribution
of iron abundances shown in Fig.~\ref{kern}. 
Thus, while in other clusters, sub-populations formed out of
   materials that were enriched by AGB ejecta, in $\omega$~Cen the
   sub-populations experienced enrichment by SN products as well. 
   This, of course, does not preclude the possibility that --as in
   other clusters --some sub-populations
may have the same iron abundance and different sodium
and oxygen abundances. This is to say that iron alone may not suffice
to identify all $\omega$~Cen sub-populations, as indeed it does not
suffice for most other clusters.

In addition, we should consider that the large variations in He abundance 
present in $\omega$~Cen
could have a significant effect on the transformation from [Fe/H] into metal content,
due to the not negligible variations in the H content. A similar effect has
been noticed by Bragaglia et al.\ (2010) in the case of NGC~2808, where a
difference of about 0.07 dex in [Fe/H] was obtained between the Na- (and
likely) He-poor population (found more metal-poor), and the Na- and He-rich
population: this difference can be entirely attributed to a variation in
the H content from $X\sim 0.75$\ down to $X\sim 0.64$, which corresponds to 0.07~dex.  
Since the range of He abundances is similar in $\omega$~Cen and
NGC~2808, we should consider that the metal content of the He-rich population
is likely to be slightly overestimated with respect to that of the He-poor population
when [Fe/H] is used as a metal abundance index.

To study the behaviour of the Na-O anticorrelation in $\omega$~Cen,
we first divided our sample into three broad
groups, following Villanova et al.\ (2007): i) a metal poor one with
[Fe/H]$\leq -1.5$ (MP); ii) an 
intermediate-metallicity group with $-1.5<$[Fe/H]$\leq -1.2$ (MInt); and iii)
a metal rich (MR) group formed by stars with [Fe/H]$>-1.2$. 

The Na and O abundances for our sample of stars in the Na-O plane are shown in
the two panels of Fig.~\ref{NaO}. 
On the left panel we represent the Na and O abundance ratios relative to iron.
The two grey lines separate the
three regions on the Na-O plane occupied by the primordial (region at high O
and low Na), intermediate (region delimited by the two lines), and extreme
(region at low O and high Na delimited by the oblique line) populations as
defined by Carretta et al.\ (2009).  
Note that these three populations identified on the basis of the position on
the Na-O plane, have been defined for {\it normal} mono-metallic
GCs. Obviously, in the case 
of $\omega$~Cen it is more difficult to  
separate three groups of stars on these basis, not only because of the spread
in iron, but  
also because a significant
number of stars lies in anomalous positions with respect to the three regions
of Carretta et al.\ (2009). 
On the right panel the Na-O anticorrelation has been also represented in the
log$\epsilon$(Na)-log$\epsilon$(O) plane. 
In both panels, MP and MInt stars are plotted as open circles and red
triangles respectively, while MR stars are represented as open stars.  
We immediately note that a Na-O anticorrelation is present within almost
all $\omega$~Cen sub-populations, although with a large dispersion both in Na
and O, and with a fraction of stars with abundances [Na/Fe]$\gtrsim$0.8
occupying an apparently anomalous  
(with respect to the other GCs) position on the Na-O plane.
Both the MP and MInt groups exhibit the Na-O anticorrelation with MInt stars
having, on average, lower oxygen abundance and higher sodium. 
Apparently, no strong evidence for a Na-O anticorrelation has been found for
stars  with [Fe/H]$> -$1.2. 
Almost all the stars in this metal rich group are Na rich, and the sample
spans a large range in O. 

As previously discussed, our results confirm that $\omega$~Cen could host more
than three groups of stars with different iron. 
An inspection of Fig.~\ref{kern} shows that the metallicity distribution has 
multiple ``peaks'' at
[Fe/H]$\sim -$1.95, $-$1.75, $-$1.60, $-$1.45, and $-$1.00, and one at $-0.75$.
It is impossible to assess whether these peaks can define six
distinct stellar groups, each with a different iron abundance.
In any case, our purpose 
here is simply to analyze the behaviour of the Na-O anticorrelation as iron
increases. 
To further investigate the Na-O anticorrelation properties in $\omega$~Cen
sub-populations, in Fig.~\ref{nao_new}, we arbitrarily isolated six groups of
stars around the iron peaks mentioned above. 
 The exact  interval of [Fe/H] corresponding to each group is quoted in the
 upper panels, where we show the Na-O anticorrelation for stars in each
 selected group. 
This procedure by no means guarantees that sub-populations have been
correctly isolated from each other, and contamination across adjacent
bins must certainly exist given the $\sim 0.10$ dex 1$\sigma$ error in
[Fe/H]. This diffusion of stars in and out each bin has to be taken in
full account when discussing the chemical abundances of stars within
each of the [Fe/H] bins discussed below.
In all the panels the selected stars in the various Fe bins are represented 
as red points, and the two grey lines are as in Fig.~\ref{NaO}.
Lower panels of Fig.~\ref{nao_new} show the position in the {\it V} vs.\ {\it B$-$V} CMD of stars in the corresponding iron interval.
In order to better follow the trend of the Na-O anticorrelation as iron 
increases,
we plot in each panel a fiducial line, drawn by hand, tracing
the anticorrelation for the population with $-$1.65$\leq$[Fe/H]$<-$1.50 (third panel from the left).
The Na-O anticorrelation is present over a wide metallicity range.
It is interesting to note that, when we compare the Na-O
    anticorrlation in each panel agains the fiducial line, the average
    Na level (abundance) increases systematically as we go from the
    metal-poor to the metal-rich populations.

The fraction of stars with low and intermediate O content also increases
with metallicity.
The Na-O anticorrelation disappears for stars with [Fe/H]$\gtrsim -$1.05, where most of the stars occupy
a quite anomalous portion of the Na-O plane, at high Na abundance and with large spread in O. 
In green and blue symbols we highlight the position of stars with an anomalous
position on the Na-O plane with respect to the bulk of stars in their
arbitrary metallicity bins (possibly due to errors in [Fe/H] measurements
which causes the migration of stars in adjacent bins). The corresponding
position of these stars on the CMD is also shown in the lower panels with the
same colors as in the upper panels. 

We now discuss the properties of stars in each bin in more detail.

\smallskip\par\noindent{\bf Bin 1, [Fe/H] $<-1.9$}, the most metal poor
component. This very metal poor group does not appear in the Sollima et
al.\ (2005b) 
photometric metallicity distribution, most likely because the color is
saturated below [Fe/H]$\sim -1.65$.
However, some hint of a peak in the metallicity distribution among values of
$-$1.8 and $-$1.9, is present in the Str\"omgren photometry-based study by
Calamida et al.\ (2009). 
This low-iron peak in Fig.~\ref{kern} may or may not be an extension of the main
peak at [Fe/H]=$-$1.75, reaching more than $\sim 2\sigma$ away from
it. We assume this peak to be a real, very metal poor component, with
most stars showing {\it primordial} Na and O abundances, i.e., high
oxygen and low sodium. Two or three O-poor stars may well belong to
the next bin, so it appears likely that this group is chemically
homogeneous as far as iron and oxygen are concerned, possibly with a
spread only in sodium.

\smallskip\par\noindent {\bf Bin 2, $- 1.90\leq$[Fe/H] $< -1.65$}, this is
the main iron-poor group, and these red giants are assumed to be the progeny of
the rMS, i.e., the main {\it first generation}. 
However, many stars have high sodium and low oxygen, and
therefore, being made of materials that have been highly processed
through the CNO-cycle and proton-capture reactions, cannot belong to
the first generation. We consider it more likely that this group
actually consists of a mixture of first and second generation stars,
i.e., the progeny of both the rMS and the bMS stars,
with the latter ones having possibly migrated to this [Fe/H] bin from the next
adjacent bin whose width is only marginally larger that our internal error in
[Fe/H]. Still, as in the previous bin, sodium abundances extend
significantly above the limit for the field stars.

\smallskip\par\noindent {\bf Bin 3, $-1.65\leq$[Fe/H] $<-1.50$}, this
[Fe/H] peak should correspond to the bMS, i.e., the main helium-rich
second generation component.  Indeed, most stars are oxygen poor and
sodium rich, as expected for the high helium sub-population. Just a
few stars appear to have the {\it primordial} composition (oxygen
rich, sodium poor), and it is likely that these may have migrated from Bin
2. Given the 
narrowness of this bin, it is expected that there will be both 
      contamination from and loss to adjacent bins, and we conclude that its
      main component, 
oxygen poor and sodium rich, may be homogeneous in these
elements. However, as shown in Fig.~\ref{spectra}, this bin includes stars with
indistinguishable [Fe/H], but different sodium and oxygen,
indicating that there must be a component that has formed out of
CNO-processed material not further enriched in iron.

\smallskip\par\noindent {\bf Bin 4, $-1.50\leq$[Fe/H] $<-1.05$}, this is
the broadest of all bins, and likely consists of 2 or 3 minor
sub-populations. The distinct oxygen-poor, sodium-rich group 
could be a
contaminant from the previous bin (and being made of bMS progeny),
and it may be homogeneous in these elements. One group with
intermediate oxygen and very high sodium is clearly distinct from the
others, and looks homogeneous. More puzzling is the group with
near-pristine oxygen and sodium, unlikely to represent objects
migrated from Bin 2 because of errors in [Fe/H]. Some stars in this bin likely
represent the RGB progeny of the so-called anomalous Main Sequence
(MSa). Indeed, stars highlighted in blue and green, which lie well outside the  
conventional Na-O anticorrelation (dashed line), also occupy a different region 
on the CMD with respect to the bulk of stars present in this bin and represented in red.
These stars appear to be distributed between the RGB sequence defined by the red
stars in this bin and the RGBa of Pancino et al.\ (2000).  
Note that the group with intermediate oxygen and very high sodium, 
represented by green symbols, is clearly distinct from the
others, and looks homogeneous. 
Overall, this bin includes at least three fairly distinct groups, each
being possibly chemically homogeneous.

\smallskip\par\noindent {\bf Bin 5, $-1.05\leq$[Fe/H] $<-0.90$}, this is a
very narrow bin, and is dominated by the very high sodium,
intermediate-oxygen stars already found in the previous bin, together
with three stars with apparently pristine oxygen and sodium to iron
ratios. Being so narrow in [Fe/H], it is likely that this bin has lost
stars to the two adjacent bins, where indeed this very sodium-rich
population is also found. The three stars with low Na and high O, represented
in blue, may have diffused from the previous iron bin, and also their position
on the CMD appears to be slightly on the blue side of the RGBa, rather than
belong to the RGBa itself. 

\smallskip\par\noindent {\bf Bin 6, [Fe/H] $\geq-0.90$}, photometrically,
this iron-rich group looks essentially identical to the main component
in the previous group (see also lower panels in Fig.~\ref{nao_new}), and like the previous
group, it represents the RGBa progeny of MSa stars. One may argue that the
very 
high-sodium stars in the last three bins are essentially identical.
However, there appears to be a possible trend of increasing [O/Fe] with
increasing [Fe/H], i.e.\ a systematic increase of [O/Fe] from Bin 4 to
Bin 6. 
Alternatively, this extreme (MSa) population may actually exhibit a
spread in both [Fe/H] and [O/Fe], with oxygen increasing more rapidly
than iron, quite an odd behaviour if most of iron comes from Type Ia
supernovae and most oxygen comes from core collapse supernovae. However,
in the high resolution CMD from WFC3 data of Bellini et al.\ (2010), the MSa
is clearly spread into two sequences, which evolve 
into two distinct SGBs. Therefore, the [O/Fe] differences among the very high
sodium stars in Bins 4-6 may well be real, and the result of having the two
components of the MSa population being spread in various proportions among
these [Fe/H] bins. 
Again, this group of sodium-rich stars is completely out of the 
conventional sodium-oxygen anticorrelation.

This analysis of the oxygen and sodium abundances in the various iron
bins can be summarized as follows:

\begin{enumerate} 
\item None of the iron bins appears to isolate
one single sub-population, except perhaps Bin 6. However, this is
possibly due to diffusion from adjacent bins,

\item The data are consistent with each [Fe/H]
bin containing 1 to 3 distinct sub-populations, each being homogeneous
in iron and oxygen abundance, while possibly exhibiting a modest
spread in sodium.

\item In the three most metal rich bins there is a
sodium rich component that does not follow the conventional Na-O
anticorrelation, with actually a hint for a Na-O correlation (cf.\ also
Fig.~\ref{NaO}).

\item 
Analysis of individual stars shows that in some cases there is no one-to-one
correspondence between [Fe/H] and the sodium and oxygen abundances,
i.e., some stars exist that have virtually identical iron but very
different sodium and oxygen (see panels {\it c} and {\it d} in
Fig.~\ref{ONaLa} and Fig.~\ref{spectra}). This is certainly not 
unique to $\omega$~Cen, as indeed most clusters exhibit a single
iron abundance while hosting stars with different oxygen and sodium
abundances.

\end{enumerate}

\section{Discussion}

In this paper, we have presented Na, O, Ba and La abundances, in a large,
homogeneously  
analyzed sample of $\omega$~Cen RGB stars. These measurements allowed us to
study,  
with a large sample of stars, the chemical pattern of different metallicity
groups hosted by the cluster. 

From our results, the picture of the chemical enrichment history in
$\omega$~Cen  
turns out to be even more complicated than previously thought:
\begin{itemize}
\item We confirm that $\omega$~Cen exhibits large star-to-star variations in
  iron,  
with [Fe/H] ranging from $\sim -$2.0 to $\sim -$0.7 dex. The iron distribution
is  
characterized by several peaks that can clearly be associated with stellar
groups located in different  
RGB regions.
\item Barium and lanthanum increase with iron for stars with [Fe/H]$\lesssim
  -$1.5,  
while their abundance distribution is flatter for the more metal rich
populations  
([Fe/H]$\gtrsim -$1.5).
\item The Na-O anticorrelation is present among $\omega$~Cen stars (as
  first found by NDC95). The presence of stars defining a Na-O
  anticorrelation has been detected over almost all the observed
  metallicity range, but with differences in the fractions of
  primordial, intermediate and extreme population stars. The metal
  poor bin (with [Fe/H]$\lesssim -$1.9) hosts a large fraction of
  stars whose Na and O abundances resemble that of the halo field
  stars (primordial population), i.e.\ they are O rich and Na poor. As the
  metallicity increases, the percentage of the Na rich and O poor
  stars becomes larger, resulting in a Na-O anticorrelation more
  extended than the one observed in the metal poor regime. This result
  is relevant for an understanding of the formation of the different
  stellar populations in $\omega$~Cen, and similar, though much more
  complex than what found in M54 by Carretta et al.\ (2010). 
  Interesting, the presence of a Na-O anticorrelation has also been found in the two
   metallicity groups hosted in M22 (Marino et al.\ 2009; Marino et al.\ in prep.).
In $\omega$~Cen the   anticorrelation disappears for the most metal rich populations.
\end{itemize}

\begin{figure}[!ht]
   \centering
   \plotone{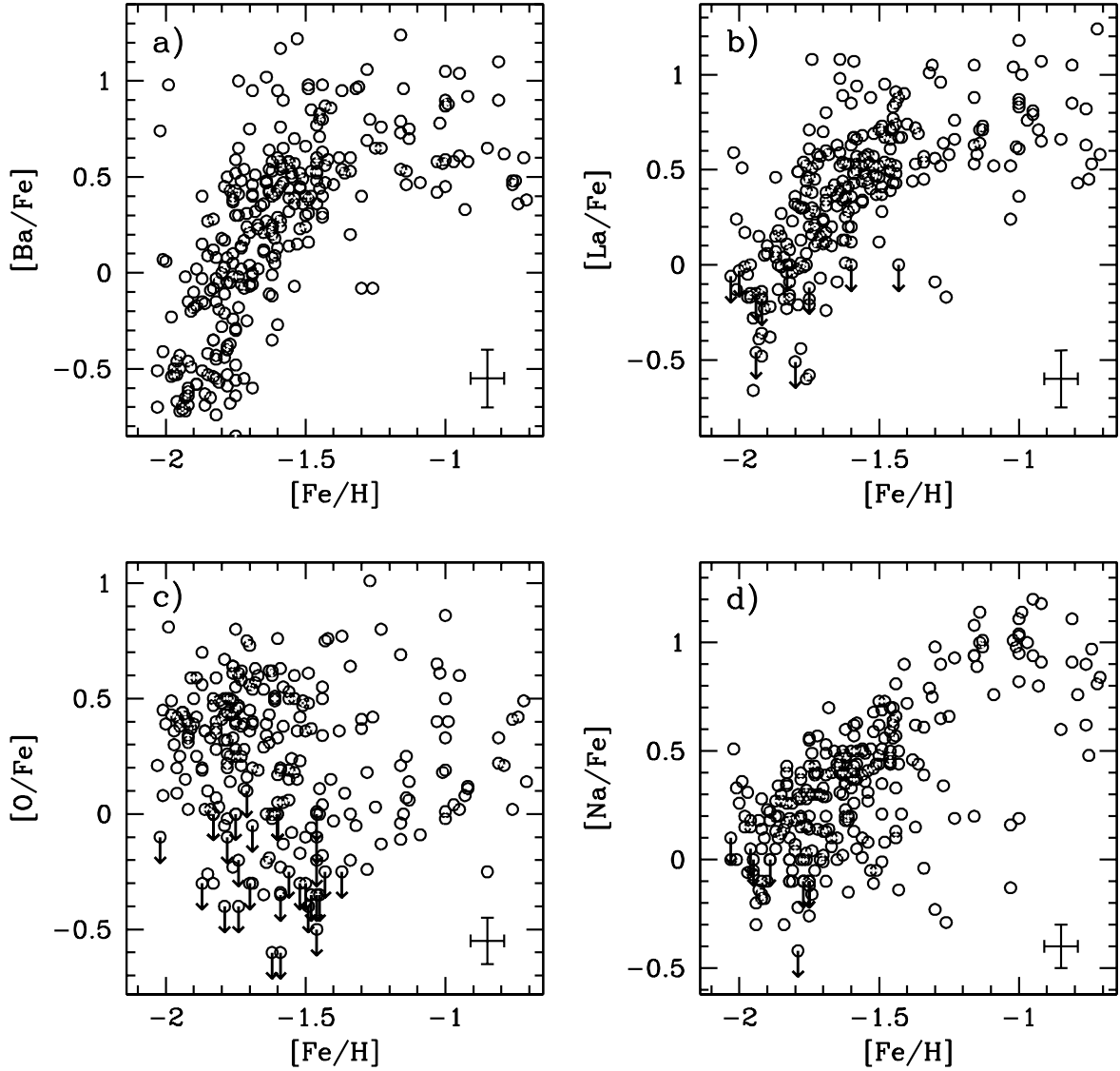}
      \caption{{\it Upper panels}: abundance ratios for the two $n$-capture elements Ba and La relative to Fe as a function of [Fe/H]. {\it Lower panels}: abundance ratios for the $p$-capture elements O and Na relative to Fe as a function of [Fe/H].} 
         \label{ONaLa}
   \end{figure}

The runs of barium and lanthanum abundances with [Fe/H] provide
further (puzzling) information on the chemical enrichment history of
$\omega$~Cen. 
Notice that such a fast raise of \spro\ increase with iron shown in
Fig.~\ref{ONaLa} is not found in other galactic  
environments, as shown in Fig.17 of J10. A rise in the \spro\ elements has
been observed in some dwarf spheroidal (see Tolstoy, Hill \& Tosi
2009). However, differently from $\omega$~Cen whose stars show enhanced
[$\alpha$/Fe] ratios, dwarf galaxies exhibit low [X/Fe] ratios for the light
and especially $\alpha$ elements (Venn et al.\ 2004; Geisler et al.\ 2007)
suggesting a significant contribution from Type Ia SNe. In addition, in dwarf
spheroidals high levels of Ba and La are reached at a much higher metallicity
than in $\omega$ Cen. 

Ba and La are produced by $n$-capture processes, either
$s$- or $r$-processes. Smith et al.\ (2000) and J10 found that Ba and La in
$\omega$~Cen metal-rich stars should be produced mainly by the \spro,\
because Eu (whose fractional contribution by \spro\ is of only $\sim$3\%,
Simmerer et al.\ 2004) abundances are roughly constant, 
independently of [Fe/H], ruling out the \rpro\ as an important
contributor. In the Sun, the \spro\ contribution is mainly due to two
components: the main component (attributed to low mass AGB 
stars of 1.5--3~M$_\odot$, Busso, Gallino \& Wasserburg 1999),
  and the weak component (attributed to massive stars, see Raiteri et
  al.\ 1993, and references therein). At solar metallicity, this last
  component mainly produces the lighter \spro\ nuclei, at most up to
  Sr and Kr (Raiteri et al.\ 1993). However, in a metal poor, and
    expecially $\alpha$-element 
  enhanced environment more neutrons are produced per iron seed
  nucleous, and heavier \spro\ elements are likely to be produced. 
Hence, the neutron seed is the $\rm {{}^{22}Ne(a, n)}{^{25}Mg}$ reaction, and
in an $\alpha$-enhanced environment more O (and Ne) are present to start with
per iron seed.    
It remains to be seen whether the models can produce elements as heavy as Ba
and La in stellar environments characterized by the chemical abundances
observed in $\omega$~Cen. 
The low abundance of Cu in [Fe/H]$\sim -1.0$ stars in $\omega$~Cen (Cunha et
al.\ 2002; Pancino et al.\ 2002) may argue against a large contribution by the
weak component, although with a higher neutron-to-iron ratio one expect the
weak component to be shifted towards heavier nuclei. 

As suggested by the referee, an additional possibility could be to allow
  for rotating massive stars. Rotational mixing may drive some protons into
  the He-burning region, allowing  for a primary contribution from the
  ne22(a,n) source, see for e.g., the preliminary models by Pignatari et
  al. (2008). Obviously more detailed models are required to test this.    

Alternatively, or in addition, the Ba and La enhancements 
      could come from the main component by the more massive AGB stars
      ($M\gtrsim 3~M_{\odot}$), (where the $\rm {{}^{22}Ne(a, n)}$ is also
      partly primary as well).  
The less massive AGB stars represent a well identified nucleosynthetic
  site to produce $s$-elements, via the $\rm ^{13}C(a,
  n)^{16}O$ reaction. An advantage of these stars in reproducing the
  chemical pattern observed in $\omega$~Cen, is that the
  neutron flux increases with iron seed with decreasing metallicity. However, 
these less massive AGB stars have the important disadvantage that they 
      would require a very long timescale for this \spro\ enrichment
      (on the order of a Gyr), which would be difficult to reconcile
      with the other relevant chemical-enrichment timescales set by
      the massive AGB stars ($\sim 3-8~M_\odot$, D'Antona \& Caloi 2004) and
      by the iron and $\alpha$-element enrichment ($>8~M_\odot$ stars).
Thus, it seems more likely to us that the \spro\ elements 
in $\omega$~Cen were produced by the same massive SN and AGB stars 
that were responsible for the enrichment of iron, $\alpha$-elements,
and helium and for establishing the Na-O anticorrelation, than
to assume that \spro\ ehnancement comes from a third set of 
producers, such as low-mass AGB stars.

\bigskip
{\it Acknowledgements} We warmly thank the anonymous referee whose
suggestions helped to improve the paper. The authors acknowledge support
by 
PRIN2007 (prot.\ n.\ 20075TP5K9), and by ASI under the program ASI-INAF
I/016/07/0.   A.B. acknowledges support by the CA.RI.PA.RO. foundation, and by
the STScI under the "2008 Graduate Research Assistantship" program. 

{}


   \begin{figure*}[!ht]
   \centering
\plotone{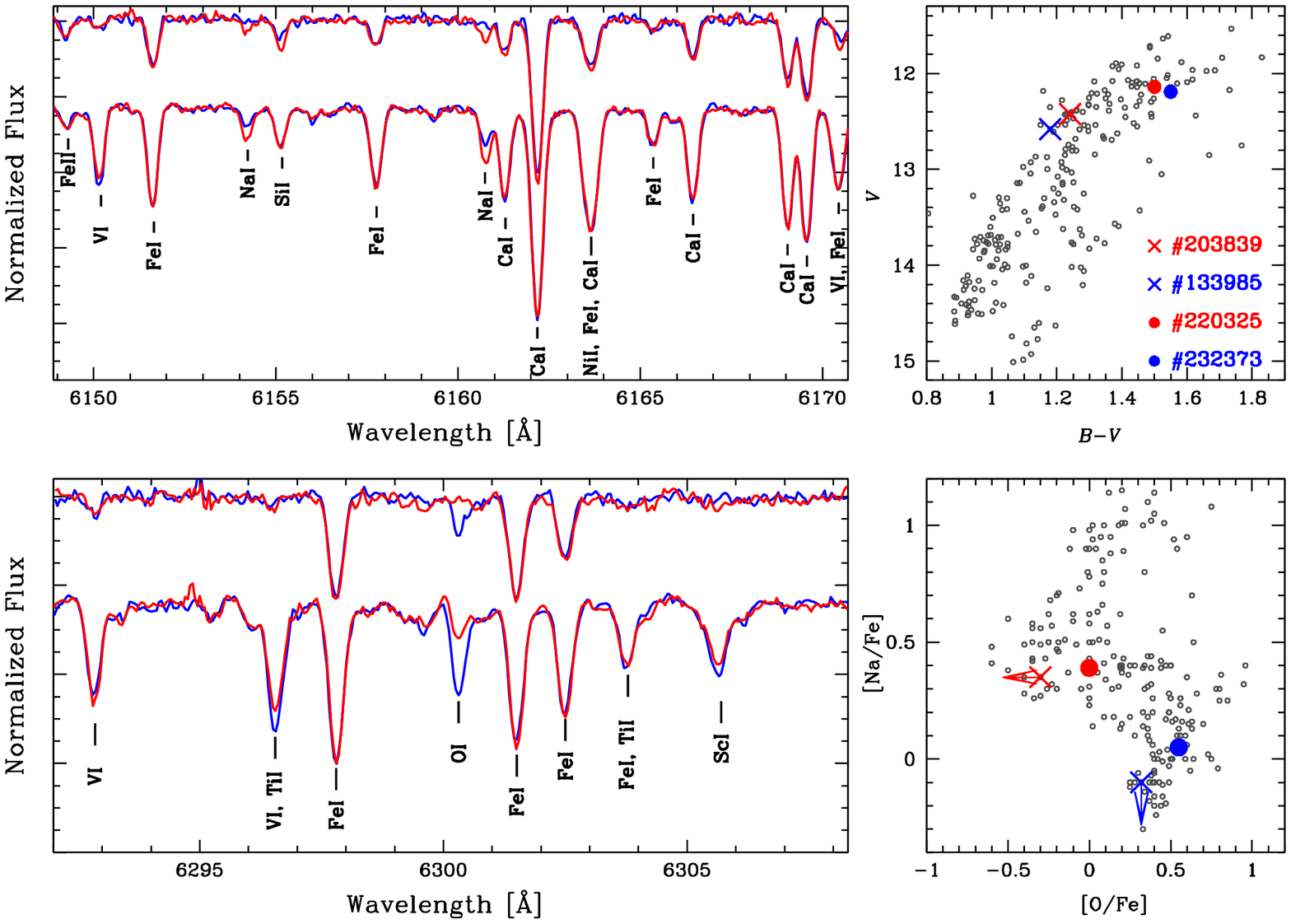}
      \caption{Spectra around Na (upper left panel) and O spectral lines (lower left panel) for two pairs of stars with similar atmospheric parameters: \#203839 [\teff=4371 K, \logg=1.08, [A/H]=$-$1.83, \vmicro=1.73 km/s] and \#133985 [\teff=4391 K, \logg=1.16, [A/H]=$-$1.79, \vmicro=1.60 km/s] are shown in the upper spectra, and \#220325 [\teff=4052 K, \logg=0.73, [A/H]=$-$1.61, \vmicro=1.90 km/s] and \#232373 [\teff=4093 K, \logg=0.79, [A/H]=$-$1.58, \vmicro=1.96 km/s] in the lower ones. The position of the two pairs of stars on the CMD and on the Na-O plane is shown on the right panels.}
         \label{spectra}
   \end{figure*}

\begin{figure}[!ht]
\centering
\plotone{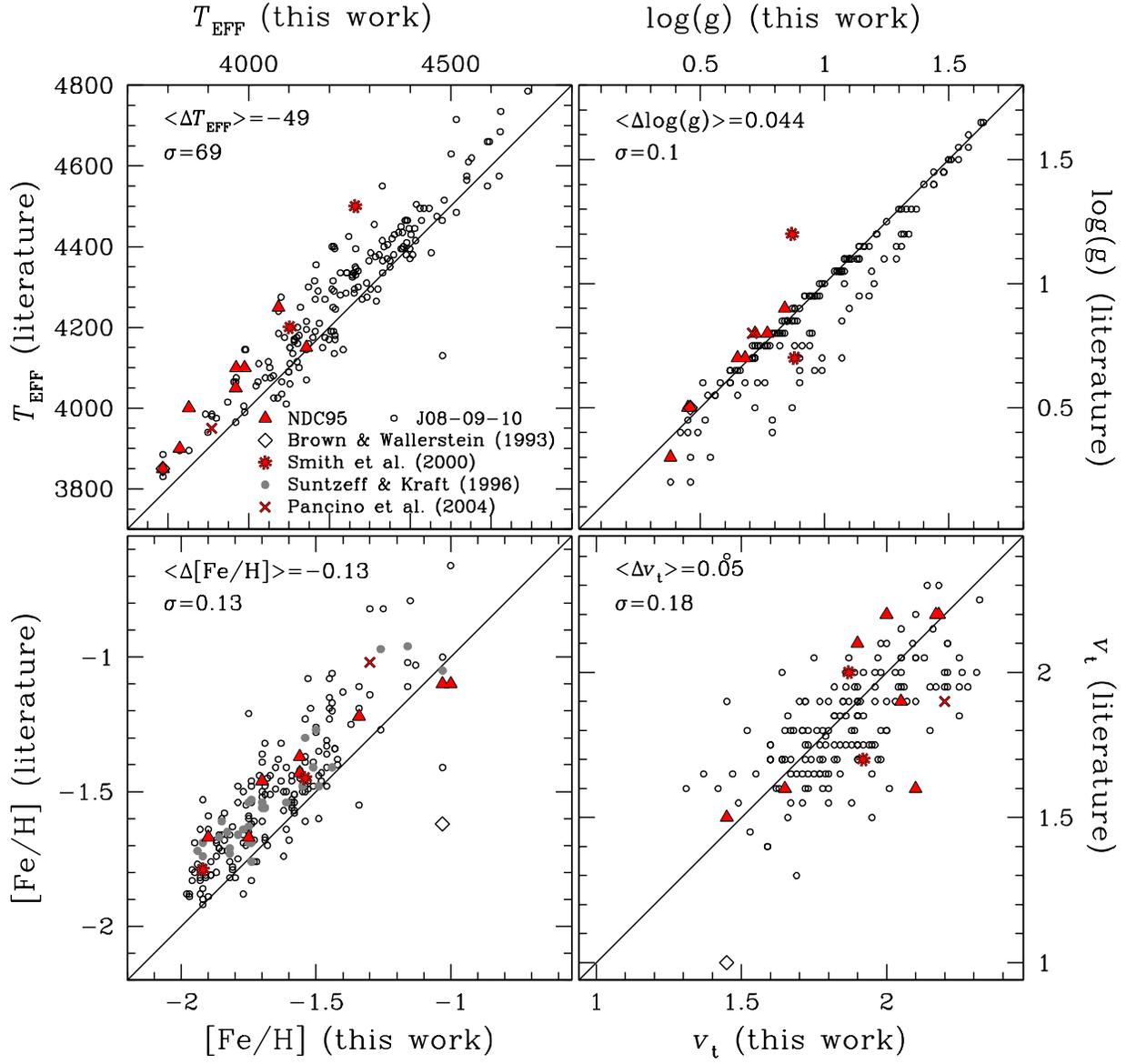}
\caption{Comparison of the model atmosphere parameters adopted in this study vs.\ those available in the literature. A straight line indicates perfect agreement in all panels. Different symbols refer to data retrieved from different studies.}
\label{bellini}
\end{figure}

\begin{figure}[!ht]
\centering
\plotone{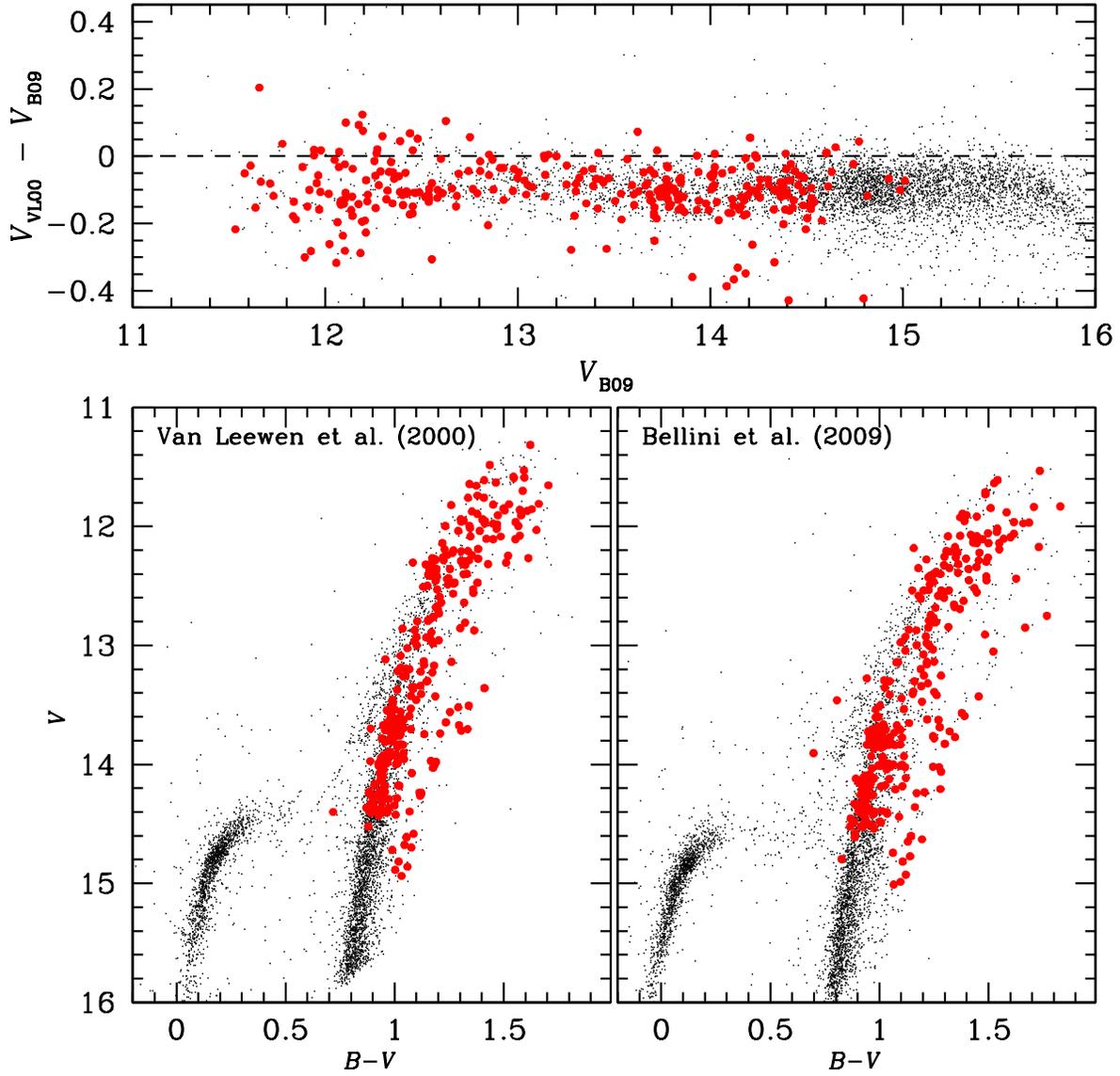}
\caption{{\it Upper Panel}: Comparison of the $V$ magnitudes retrieved from Bellini et al.\ (2009) and from Van Leween et al.\ (2000). {\it Lower Panels}: $V$-$(B-V)$ CMDs constructed from the two different photometric catalogs.
In all the panels the red points represent our spectroscopic sample.}
\label{fotometrie}
\end{figure}

\begin{figure}[!ht]
\centering
\plotone{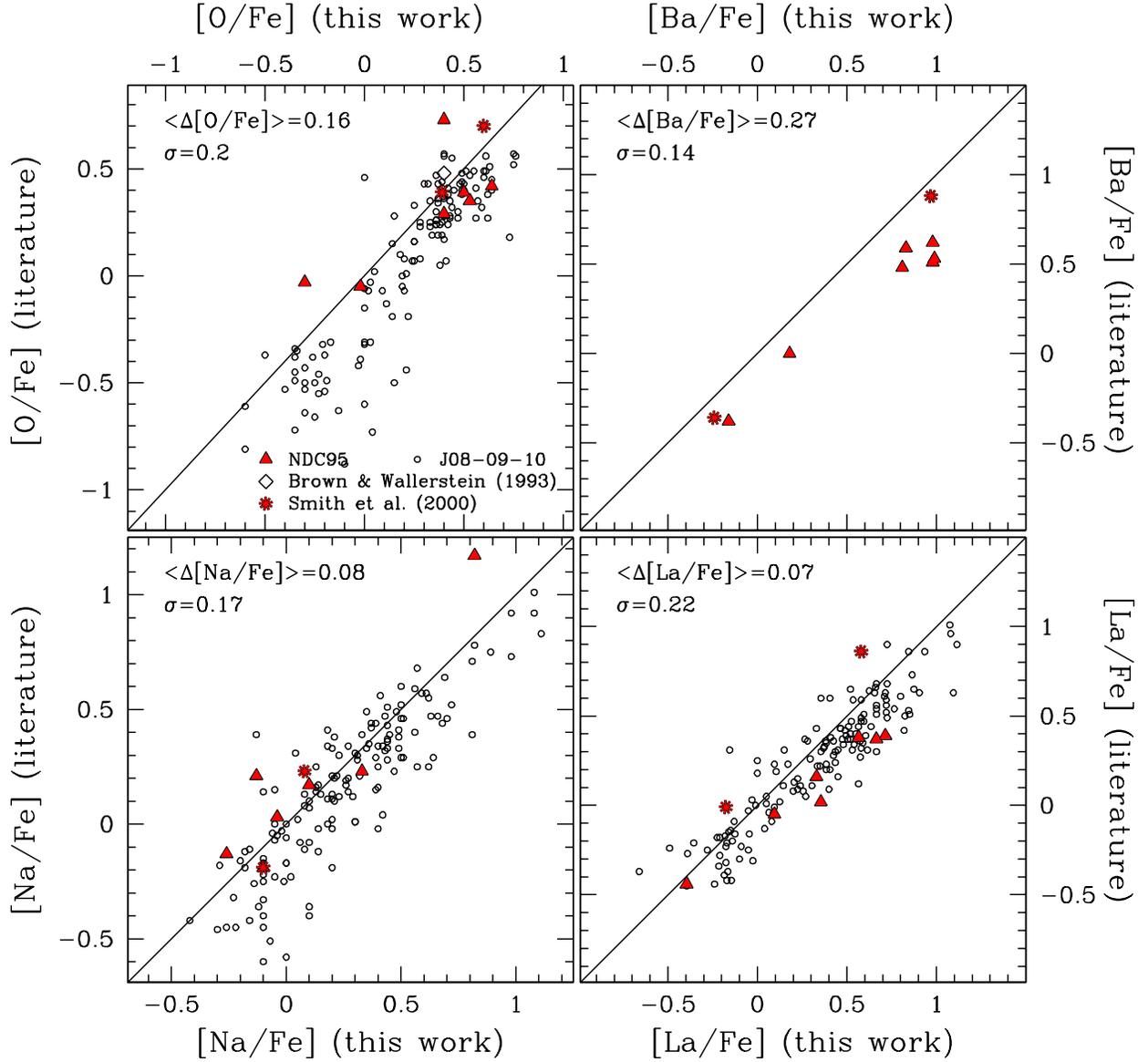}
\caption{Comparison of the abundance ratios obtained in this study vs.\ those available in the literature. A straight line indicates perfect agreement in all panels. Different symbols refer to data retrieved from different studies.}
\label{confrontoEL}
\end{figure}

   \begin{figure*}[!ht]
   \centering
\epsscale{0.95} 
   \plotone{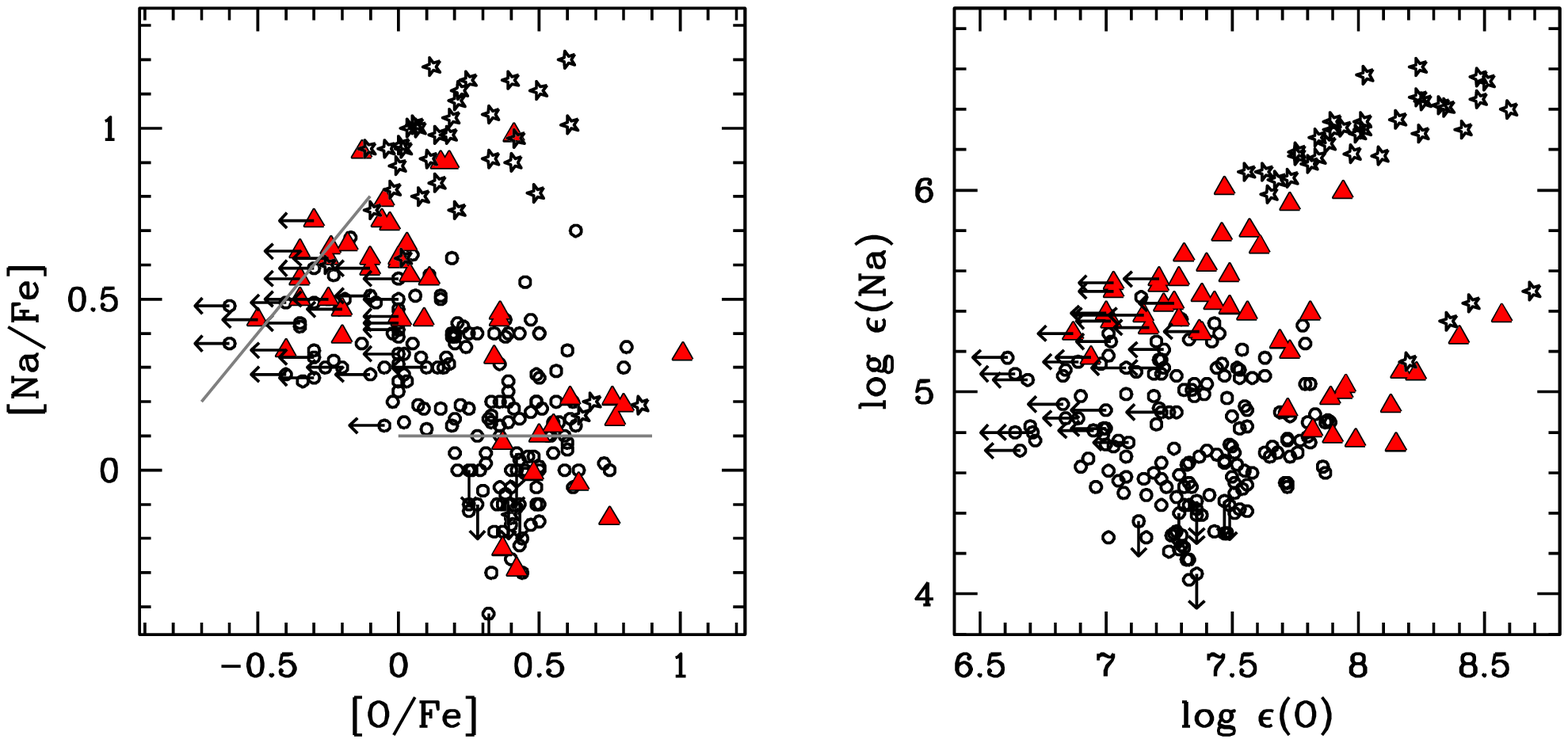}    
\vskip 35pt
      \caption{
Na-O anticorrelation for 253 stars. We have used open
        circles, filled red triangles and open stars to represent stars with
        [\rm Fe/H]$\leq -$1.50 (MP stars), $-$1.50$<$[\rm
          Fe/H]$\leq$1.20 (MInt stars), and [Fe/H]$> -$1.20 (MR
          stars) respectively.} 
         \label{NaO}
   \end{figure*}

\begin{figure*}[!ht]
\centering
\epsscale{1} 
\plotone{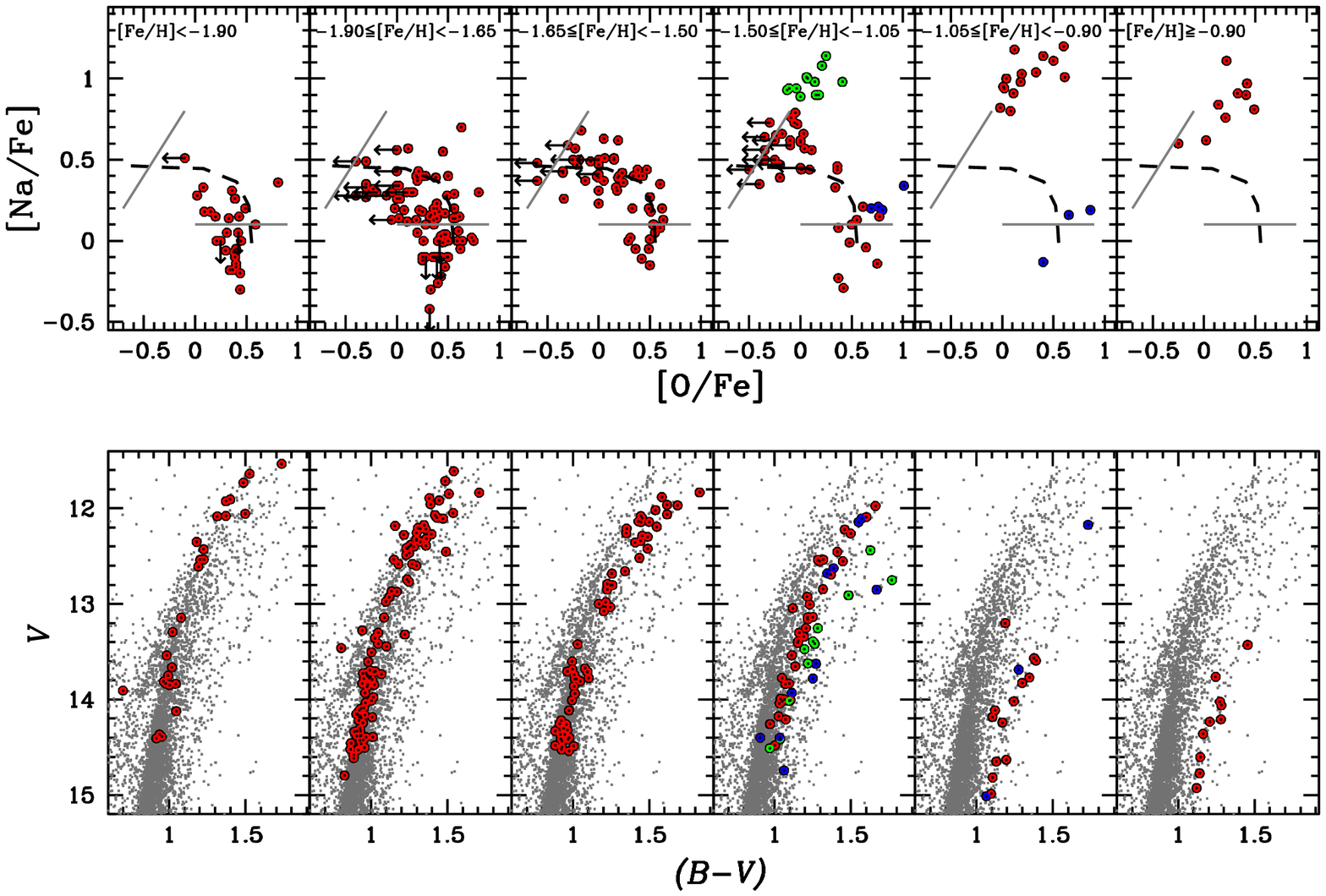}
\vskip 35pt
\caption{ {\it Lower panels}:
$V$ vs.\ $(B-V)$ CMD from WFI photometry (Bellini et al.\ 2009).
In each  panel, the red dots show stars which have been selected
according to their metal content (as indicated in the inset of the corresponding
upper panel).
{\it Upper panels}: estimated [O/Fe] and [Na/Fe] for the stars of each
metallicity group.
A fiducial, traced by hand on the Na-O anticorrelation for stars with
$-$1.65$\leq$[\rm Fe/H]$<-$1.50, has been over-imposed to each Na-O plane.
In the fourth and fifth panels, the green and blue dots highlight the position
in the CMD of the stars with extreme values in the Na-O anticorrelation, as
represented in the corresponding upper panels.}
\label{nao_new}
\end{figure*}

%

\clearpage
\begin{center}
\begin{deluxetable}{c c c c c c c c c r r r}
\tabletypesize{\footnotesize}
\tablewidth{0pt}
\tablecaption{Coordinates RA and DEC, $V$ magnitude, adopted atmospheric parameters and measured abundance ratios for the 300 stars in $\omega$~Cen. The complete table is available only in electronic form at CDS.\label{list}}
\tablecolumns{12}

\tablehead{
\colhead{ID\tablenotemark{a}}  &
\colhead{RA}                   &
\colhead{DEC}                  &
\colhead{$V$}                  &
\colhead{\teff\ [K]}           &
\colhead{\logg}                &
\colhead{\vmicro\ [km/s]}      &
\colhead{[Fe/H]}               &
\colhead{[O/Fe]}               &
\colhead{[Na/Fe]}              &
\colhead{[Ba/Fe]}              &
\colhead{[La/Fe]}               
}
\startdata
109369 & 201.818386 &$-$47.549133 & 13.044  & 4544&  1.44 &  1.70 & $-$1.43&        0.75 &  $-$0.14 &   0.56 &     0.00\\   
112383 & 201.834424 &$-$47.534160 & 12.850  & 4359&  1.24 &  1.65 & $-$1.62&        0.62 &     0.20 &   0.40 &     0.36\\   
114842 & 201.817052 &$-$47.522465 & 12.927  & 4415&  1.31 &  1.52 & $-$1.90&        0.39 &     0.23 &$-$0.10 &     0.06\\   
116394 & 201.846235 &$-$47.515000 & 13.777  & 4663&  1.80 &  1.35 & $-$1.71&        0.10 &     0.30 &   0.24 &  $-$0.07\\   
119799 & 201.830431 &$-$47.499341 & 14.023  & 4293&  1.67 &  1.40 & $-$0.71&        0.14 &     0.84 &   0.38 &     0.58\\   
120847 & 201.807438 &$-$47.494576 & 13.034  & 4343&  1.31 &  1.55 & $-$1.51&        0.47 &     0.44 &   0.49 &     0.47\\   
121048 & 201.834524 &$-$47.493559 & 13.781  & 4394&  1.64 &  1.48 & $-$1.16&        0.69 &     0.20 &   0.54 &     0.53\\   
126851 & 201.854789 &$-$47.466319 & 14.061  & 4274&  1.67 &  1.55 & $-$0.72&        0.49 &     0.81 &   0.60 &     1.24\\   
131171 & 201.849422 &$-$47.445934 & 13.760  & 4648&  1.78 &  1.60 & $-$1.85&     \nodata &  \nodata &$-$0.42 &  \nodata\\  
131256 & 201.823805 &$-$47.445611 & 13.407  & 4428&  1.51 &  1.53 & $-$1.34&     $-$0.20 &     0.39 &   0.53 &     0.57\\   
\hline
\enddata
\tablenotetext{a}{Identification from the photometric catalog of B09.}
\end{deluxetable}

\end{center}

%

\end{document}